\documentclass[12pt]{article}
\usepackage[centertags]{amsmath}
\usepackage{amsfonts}
\usepackage{amssymb}
\usepackage{amsthm}
\usepackage{newlfont}

\numberwithin{equation}{section}

\newtheorem{theorem}{Theorem}

\begin{document}
\title{ Eigenvalues of the generalized Laplacian and the G-dynamics of type I}
\author{Jack Whongius\thanks{School of Mathematical Sciences,~ Xiamen
University,~ 361005,  China.  Email:  fmsswangius@stu.xmu.edu.cn.}}

\maketitle

\par
\begin{abstract}
  In this paper, we consider the generalized Laplace operator equipped with the G-dynamics operator of type I, the Dirichlet and Neumann eigenvalue problems are extended to associate with the G-dynamics of type I, it is proved that the G-dynamics of type I satisfies an integral identity. The G-dynamics of type II for generalized Laplacian is studied as well. Using the general method related to Dirichlet eigenvalue problem, an estimate analysis for the generalized Laplacian with some conditions is made.

\end{abstract}



\section{Introduction}

\subsection{Laplace operator}

Laplace operator $ \Delta $, also called Laplacian as a differential operator in $ \mathbb{R}^n$ is defined by  $\Delta =\sum\limits_{i=1}^{n}{{{\partial }^{2}}/\partial x_{i}^{2}}$,
where $x_{1}, \dots, x_{n}$ are coordinates in $\mathbb{R}^n$. The Laplacian is an elliptic second order differential operator [1-6]. The Laplace operator also plays an important role such as Laplace equation, Laplace-Beltrami equation $\Delta =-\frac{1}{\sqrt g}\frac \partial{\partial x^{i}}\left ( \sqrt g g^{ij} \frac{\partial}{\partial x^{j}}\right)$  with the Riemannian metric, harmonic function, harmonic form, etc [15-17]. The Laplacian also appears in the famous quantum equation--Schr\"{o}dinger equation
related to the quantum spectrum.
When it comes to the space of exterior differential forms on $M$, the Laplacian shows us the form
$\Delta=(d+d^{*})^{2} = dd^{*} + d^{*} d$, where $d$ is the operator of exterior differentiation and $d^{*}$ is the operator formally adjoint to $d$ [1,2,3].

Laplace operator $\Delta$ corresponds to the eigenvalue equation
\begin{equation}\label{b2}
  -\Delta \varphi=\lambda \varphi,~~x\in \Omega,
\end{equation}
also called Helmholtz equation which is exactly the eigenvalue problem of Laplace operator. We know that $\Delta$ is self adjoint. Therefore, under given boundary value conditions, such as Dirichlet boundary conditions and Neumann boundary conditions, the spectrum of this operator is complete and countable. We arrange it from small to large $0<{{\lambda }_{1}}<{{\lambda }_{2}}\le {{\lambda }_{3}}\le \cdots$. The value of its first non-zero eigenvalue ${{\lambda }_{1}}$ is called the first eigenvalue. When studying the steady state of the equation, that is, the large time behavior of the equation solution, it is of great significance [8,11-13].

For Helmholtz equation \eqref{b2},  let $\Omega\subset \mathbb{R}^n$ be a bounded open region, then there is an unbounded sequence $\left\{\lambda_k\right\}$ with monotonic rise satisfying: $0<{{\lambda }_{1}}\le {{\lambda }_{2}}\le \cdots$ , ${{\lambda }_{k}}=+\infty $ as $k\to \infty$ and a group of Hilbert base $\left\{\varphi_k\right\}$ of ${L}^2(\Omega)$ such that $-\Delta {{\varphi }_{k}}={{\lambda }_{k}}{{\varphi }_{k}}$.

The eigenvalue problem of Laplace operator is one of the core research directions in many mathematical fields, for instance, geometric analysis, partial differential equations, variational calculus, mathematical physics, etc. At present, there are still many open problems that have not been solved [12].

The eigenvalue problem of Laplace operator is described by the formula \eqref{b2},
and the most common boundary conditions are shown by the following:
\begin{description}
  \item[Dirichlet:] $\varphi \left| _{\partial \Omega } \right.=0$.
  \item[Neumann:] $\frac{\partial \varphi }{\partial \boldsymbol {n}}\left( x \right)=0,~x\in \partial \Omega.$
\end{description}respectively.
Notice that in the Neumann case above, where $\frac{\partial \varphi }{\partial \boldsymbol {n}}\left( x \right)=0$ is the directional derivative of $\varphi$ in the direction of the outward pointing surface normal $\boldsymbol {n}$ of the surface element $dS$, that is,
${\displaystyle {\partial \varphi  \over \partial \boldsymbol {n} }=\nabla \varphi \cdot \boldsymbol {n} }$.

\subsection{Neumann eigenvalue problem}

Let $\Omega \subset {{\mathbb{R}}^{n}}$ be a bounded domain with a piecewise smooth boundary $\partial \Omega$. If there exists a Neumann eigenfunction $\varphi \in {{C}^{2}}\left( \Omega  \right)\cap {{C}^{0}}\left( \overline{\Omega } \right)$ such that Neumann boundary conditions holds below [1]:
\begin{equation}\label{a1}
  \left\{ \begin{matrix}
   -\Delta \varphi \left( x \right)=\mu \left( x \right)\varphi \left( x \right),~x\in \Omega   \\
   \frac{\partial \varphi }{\partial \boldsymbol {n}}\left( x \right)=0,~x\in \partial \Omega .  \\
\end{matrix} \right.
\end{equation}then $\mu$ is a Neumann eigenvalue on $\Omega$,
where $\frac{\partial \varphi }{\partial \boldsymbol {n}}\left( x \right)$ is the normal derivative of $\varphi$.   The Neumann Laplacian has a discrete spectrum of infinitely many non-negative eigenvalues with no finite accumulation point:
$0=\mu_{1}(\Omega)\leq\mu_{2}(\Omega)\leq\cdots$, ~
${{\mu }_{k}}\rightarrow\infty$ as $k\rightarrow\infty $.

Furthermore, Neumann eigenvalue problem of weighted Laplace operator refers to the following problems:
\[\left\{ \begin{matrix}
   -\Delta \varphi \left( x \right)=\mu \rho \left( x \right)\varphi \left( x \right),~x\in \Omega   \\
   \frac{\partial \varphi }{\partial \boldsymbol {n}}\left( x \right)=0,~x\in \partial \Omega .  \\
\end{matrix} \right.\]
 This kind of problem is also an important research content in spectral theory and applied mathematics.

\subsection{Dirichlet eigenvalue problem}

Consider Dirichlet eigenvalue problem (DEP)[14,18]
\begin{equation}\label{a2}
  \left\{ \begin{matrix}
   -\Delta \varphi =\lambda \varphi ,\begin{matrix}
   {} & in  \\
\end{matrix}~\Omega,   \\
   \varphi \left| _{\partial \Omega } \right.=0.  \\
\end{matrix} \right.
\end{equation}
As we all know, DEP on $\Omega$ has discrete eigenvalues. It can be shown, using the spectral theorem for compact self-adjoint operators that the eigenspaces are finite-dimensional and that the Dirichlet eigenvalues $\lambda$ are real, positive, and have no limit point. Thusly, the Dirichlet Laplacian has a discrete spectrum of infinitely many positive eigenvalues with no finite accumulation point:
 $0<{{\lambda }_{1}}\leq{{\lambda }_{2}}\le \cdots $,~~
 ${{\lambda }_{k}}\rightarrow\infty$ as $k\rightarrow\infty $.

For large values of $k$, if $\Omega \subset {{\mathbb{R}}^{n}}$, H. Weyl [9,10] proved
$${{\lambda }_{k}}\approx \frac{4{{\pi }^{2}}{{k}^{2/n}}}{{{\left( {{C}_{n}}\left| \Omega  \right| \right)}^{2/n}}},$$
where $\left| \Omega  \right|$ and ${{C}_{n}}={{\pi }^{n/2}}\Gamma \left( n/2+1 \right)$ are, respectively, the volumes of $\Omega$ and of the unit ball in ${{\mathbb{R}}^{n}}$ [2,18]. So ${{\lambda }_{k}}\left( \Omega  \right)$ represents the $k$-th characteristic value of DEP on $\Omega$. The term the $k$-th Dirichlet characteristic function satisfies $\lambda ={{\lambda }_{k}}$.
The boundary value problem is the Dirichlet problem for the Helmholtz equation, and so $\lambda$ is known as a Dirichlet eigenvalue for $\Omega$.

\subsection{The G-dynamics of type I}
The quantum covariant Poisson bracket
$\left[ \hat{f},\hat{g} \right]={{\left[ \hat{f},\hat{g} \right]}_{QPB}}+G\left( s,\hat{f},\hat{g} \right)$ as a natural extension of the quantum Poisson brackets $\left[ \hat{f},\hat{g} \right]_{QPB}=\hat{f}\hat{g}-\hat{g}\hat{f}$ is the generalized geometric commutator of two operators $\hat{f},~\hat{g}$,  where  $$G\left(s, \hat{f},\hat{g} \right)=\hat{f}{{\left[ s,\hat{g} \right]}_{QPB}}-\hat{g}{{\left[ s,\hat{f} \right]}_{QPB}},$$ is the quantum geometric bracket, and in the formula, the geometric scalar potential function or  geometric structure function $s$ only represents the attributes of the manifold space itself, general speaking, various manifolds has different geometric scalar potential function $s$,  it's chosen a form of logarithmic function generally. From the perspective of research, it can take any functional form.
Based on the quantum covariant Poisson bracket which consists of three parts, and two parts are the contribution of the quantum geometric bracket,  it naturally has given the quantum covariant Hamiltonian system, then
the covariant dynamics $${\mathcal{D}\hat{f}}/{dt}=-i\left[ \hat{f},\hat{H} \right]/\hbar,$$ follows, where ${{i}^{2}}=-1$, and an observable $\hat{f}$, Hamiltonian operator $\hat{H}$.
As the formulation of the quantum covariant Poisson bracket $\left[ \cdot,\cdot \right]$ shows, the covariant dynamics can be mainly divided into two parts:  the generalized Heisenberg equation and the G-dynamics. More specifically,
the generalized Heisenberg equation is described by:
$$d\hat{f}/dt=-i{{\left[ \hat{f},\hat{H} \right]}_{QPB}}/\hbar +i\hat{H}{{\left[ s,\hat{f} \right]}_{QPB}}/\hbar .$$
      where an observable $\hat{f}$ is a physical quantity that can be measured, $\hat{H}$ is the Hamiltonian operator, and $s$ is the geometric scalar
potential function satisfying the formula  $$\hat{w}=-i{{\left[ s,\hat{H} \right]}_{QPB}}/\hbar, $$ defined as the G-dynamics.
According to the studies of the G-dynamics, it depicts the quantum rotation of the manifolds space in a certainty quantum system, its eigenvalues describe the frequency spectrum.  Note that the geometric potential function $s$ generally exists independently of $\hat{f}$ and $\hat{H}$, in other words, geometric potential function $s$ have nothing to do with $\hat{f}$ and $\hat{H}$, it's only determined by the intrinsic structure and properties of the manifolds space.
So far as it reaches, there has three different kinds of one-dimensional G-dynamics as a quantum operator found as follows [7]:
\begin{description}
  \item[Type I:] ${{{\hat{w}}}^{\left( cl \right)}}={{b}_{c}}\left( 2u d/dx+u_{x}\right)={{b}_{c}}\hat{Q}$.
  \item[Type II:] ${{{\hat{w}}}^{\left( ri \right)}}={{\hat{w}}^{\left( cl \right)}}+{{w}^{\left( s \right)}}={{b}_{c}}\left( 2ud/dx+{{u}_{x}}+2{{u}^{2}} \right)$.

  \item[Type III:] ${{w}^{\left( s \right)}}=2{{b}_{c}}{{u}^{2}}=-i\hbar {{u}^{2}}/m$.
\end{description}
where line curvature $u=ds/dx$ and $u_{x}=du/dx$ have been used, and $b_{c}=-i\hbar/2m$, the geometric differential operator $\hat{Q}={{\hat{w}}^{\left( cl \right)}}/{{b}_{c}}
$ is the curvature operator satisfying $\hat{Q}{u}^{-1/2}=0$. Note that type I, type II, type III are respectively the self-adjoint operator,  non-self-adjoint operator and anti-self-adjoint operator, these three type given by the formula of the G-dynamics using the different Hamiltonian operator are the supplemental terms to the classical Schr\"{o}dinger equation, it can describe more wider quantum system.
Another application of these G-dynamics of three types is to equip with the Laplacian $\Delta$ for a further study in the given boundary conditions.

The one-dimensional G-dynamics of type I $${{\hat{w}}^{\left( cl \right)}}=-i{{\left[ s,{{\hat{H}}^{\left( cl\right)}}\right]}_{QPB}}/\hbar,$$ is derived from the classical Hamiltonian operator ${{\hat{H}}^{\left( cl \right)}}={{c}_{1}}{{{d}^{2}}}/d{{x}^{2}}+V\left( x \right)$, where ${{c}_{1}}=-{{\hbar }^{2}}/\left( 2m \right)$, and $V=V\left( x \right)$ is the potential energy.
The corresponding geometric wave equation  follows
\[i\hbar {{\hat{w}}^{\left( cl \right)}}\psi ={{\hat{H}}^{\left( \operatorname{clm} \right)}}\psi =-{{c}_{1}}\left( 2u{{\psi }_{x}}+\psi {{u}_{x}} \right),\]where the imaginary geomenergy is
${{\hat{H}}^{\left( \operatorname{clm} \right)}}=-{{c}_{1}}\left( 2u{d}/{dx}+{{u}_{x}} \right)$ in terms of the space coordinate for convenience.

The one-dimensional G-dynamics of type II with respect to ${{\hat{H}}^{\left( ri \right)}} $ is given by
$$ {{\hat{w}}^{\left( ri \right)}} =-i{{\left[ s,{{\hat{H}}^{\left( ri \right)}} \right]}_{QPB}}/\hbar,$$
where the non-Hermitian Hamiltonian operator
\begin{equation}\label{a3}
  {{\hat{H}}^{\left( ri \right)}} ={{{\hat{H}}}^{\left( cl \right)}}-{{E}^{\left( s \right)}}/2-i\hbar {{\hat{w}}^{\left( cl \right)}}={{\hat{T}}^{\left( ri \right)}}+V(x),
\end{equation}
is used, and the geometrinetic energy operator
${{\hat{T}}^{\left( ri \right)}}={{\hat{E}}^{\left( w \right)}}+{{c}_{1}}{u}^{2}$,  the one-dimensional motor operator is
\begin{equation}\label{a4}
  {{\hat{E}}^{\left( w \right)}}={{c}_{1}}{d}^{2}/d{{x}^{2}}-i\hbar {{\hat{w}}^{\left( cl \right)}}.
\end{equation}The non-Hermitian geometric Hamiltonian is ${{\hat{H}}^{\left( s \right)}}=-{{E}^{\left( s \right)}}/2-i\hbar {{\hat{w}}^{\left( cl \right)}}$ follows. Accordingly,
the G-operator as a energy operator that is given by
\begin{equation}\label{b6}
  {{\hat{H}}^{\left( gr \right)}} ={{{\hat{H}}}^{\left( cl \right)}} -i\hbar {{\hat{w}}^{\left( ri \right)}}={{\hat{H}}^{\left( ri \right)}}-{{E}^{\left( s \right)}}/2.
\end{equation}
Similarly, the geometric wave equation in terms of ${{\hat{w}}^{\left( ri \right)}}$ is given by
$i\hbar {{{\hat{w}}}^{\left( ri \right)}}\psi  ={{{\hat{H}}}^{\left( \operatorname{rim} \right)}}\psi
$, where ${{{\hat{H}}}^{\left( \operatorname{rim} \right)}}=-{{c}_{1}}\left( 2u{d}/{dx}+{{u}_{x}}+2{{u}^{2}} \right)$ and
${{{\hat{H}}}^{\left( \operatorname{rim} \right)}}\psi={{{\hat{H}}}^{\left( \operatorname{clm} \right)}}\psi +{{E}^{\left( s \right)}}\psi$.

The one-dimensional G-dynamics of type III is given by
 $${{w}^{\left( s \right)}}={{\left[ {{\hat{w}}^{\left( cl \right)}},s \right]}_{QPB}}=2{{b}_{c}}{{u}^{2}},$$ and ${{E}^{\left( s \right)}}(u)=-2{{c}_{1}}{{u^{2}}}=i\hbar{{w}^{\left( s \right)}}$ can be regarded as a potential energy.
Conclusively, there are four different kinds of the non-Hermitian Hamiltonian operator as follows:

(1) ${{\hat{H}}^{\left( ri \right)}}={{{\hat{H}}}^{\left( cl \right)}}+{{{\hat{H}}}^{\left( s \right)}}={{{\hat{E}}}^{\left( w \right)}}+V-{{E}^{\left( s \right)}}/2={{{\hat{H}}}^{\left( cl \right)}}-{{E}^{\left( s \right)}}/2-i\hbar {{\hat{w}}^{\left( cl \right)}}$.

(2) ${{\hat{E}}^{\left( w \right)}}={{c}_{1}}{d}^{2}/d{{x}^{2}}-i\hbar {{\hat{w}}^{\left( cl \right)}}$.

(3) ${{\hat{H}}^{\left( s \right)}}=-{{E}^{\left( s \right)}}/2-i\hbar {{\hat{w}}^{\left( cl \right)}}$.

(4) ${{{\hat{H}}}^{\left( gr \right)}}={{{\hat{H}}}^{\left( cl \right)}}-{{E}^{\left( s \right)}}-i\hbar {{{\hat{w}}}^{\left( cl \right)}}$.

As mentioned, their common part is the attribution of the one-dimensional G-dynamics of type I which is self-adjoint operator, it broadly brings us to a general case of quantum mechanics with the complex eigenvalues. Above four different non-Hermitian Hamiltonian can be connected to one equation by using the G-operator
\begin{align}
 {{{\hat{H}}}^{\left( gr \right)}} &  ={{{\hat{H}}}^{\left( cl \right)}}+{{{\hat{H}}}^{\left( s \right)}}-{{E}^{\left( s \right)}}/2 ={{{\hat{E}}}^{\left( w \right)}}+V-{{E}^{\left( s \right)}} \notag
\end{align}
Note that ${{E}^{\left( s \right)}}$ is geometric potential energy which plays a role like the potential energy $V$, hence, in some special cases, we can let ${{E}^{\left( s \right)}}=V$ be given for a simply discussion.
Similarly,  three similar equations for describing the quantum mechanics are respectively given by
\begin{description}
  \item[Schr\"{o}dinger equation:] $i\hbar {{\partial }_{t}}\psi ={{\hat{H}}^{\left( cl \right)}}\psi ={{c}_{1}}{{\psi }_{xx}}+\psi V\left( x \right)$,~~ (Hermitian equation)
  \item[via Type I:] $i\hbar {{\hat{w}}^{\left( cl \right)}}\psi ={{\hat{H}}^{\left( \operatorname{clm} \right)}}\psi =-{{c}_{1}}\left( 2u{{\psi }_{x}}+\psi {{u}_{x}} \right)$, ~~ ( anti-Hermitian equation)
  \item[via Type II:] $i\hbar {{\hat{w}}^{\left( ri \right)}}\psi ={{\hat{H}}^{\left( \operatorname{rim}  \right)}}\psi =-{{c}_{1}}\left( 2u{{\psi }_{x}}+\psi {{u}_{x}}+2\psi {{u}^{2}} \right)$,~~ (non-Hermitian equation)

\end{description}

The anti-Hermitian equation induced by the G-dynamics of type I as a supplement to Schr\"{o}dinger equation as a Hermitian equation makes the quantum equations more integrated, it accomplishes the quantum ranges unknown to be studied. This mixed form also happens to the non-Hermitian equation induced by the G-dynamics of type II. Their mutually combinations can leads to the second order differential equation with variable coefficients, and it reveals new results to explain new quantum phenomenon. Schr\"{o}dinger equation is a special second order differential equation, anti-Hermitian equation is a special first order differential equation, They complement one another.
Obviously, these three equations can be mutually subtracted or added up in pairs, for examples
 \[i\hbar \hat{{{D}_{t}}}^{\left( cl \right)}\psi  ={{c}_{1}}\left( {{\psi }_{xx}}+2u{{\psi }_{x}}+\psi {{u}_{x}} \right)+\psi V\left( x \right).\]\[i\hbar \left( {{\partial }_{t}}+{{{\hat{w}}}^{\left( ri\right)}} \right)\psi  ={{c}_{1}}\left( {{\psi }_{xx}}-2u{{\psi }_{x}}-\psi {{u}_{x}} -2\psi {{u}^{2}} \right)+\psi V\left( x \right).\]
where D-operator related to the G-dynamics of type I is $\hat{{{D}_{t}}}^{\left( cl \right)}={{\partial }_{t}}-{{\hat{w}}^{\left( cl \right)}}$.
 This is a natural extension of classical Schr\"{o}dinger equation, in fact, these three equations respectively describes different quantum phenomenon, but it works for a larger quantum system if we bond them together such as the example given above, to give a complete description for a given quantum system makes us get further to realize the quantum world,
this way provides abundent visions to better understand how quantum world means.

\section{Generalized Laplacian and motor operator}

\par
In this sector, we try to use generalized Laplace operator to realize how the complex eigenvalues of motor operator means, based on the known results of Laplace operator related to the Dirichlet case,
and Neumann case, etc.

Consider eigenvalues equation problem of the motor operator (MEP) (1.4)
 ${{\hat{E}}^{\left( w \right)}}={{c}_{1}}\Delta -i\hbar {{\hat{w}}^{\left( cl \right)}}$, that is ${{\hat{E}}^{\left( w \right)}}\varphi =\gamma \varphi $, where $${{\hat{w}}^{\left( cl \right)}}={{b}_{c}}\left( 2u\cdot\nabla +\nabla u \right)=-i{{\left[ s,{{\hat{H}}^{\left( cl \right)}} \right]}_{QPB}}/\hbar, $$ and here $u=\nabla s$ is the gradient of geometric scalar
potential function $s$, and the classical Hamiltonian operator here is taken the form $${{\hat{H}}^{\left( cl \right)}}={{c}_{1}}\Delta+V\left( x \right),$$where $\Delta$ is the Laplacian.  As a consequence, the motor operator (MEP) (1.4) is rewritten as
\[{{\hat{E}}^{\left( w \right)}}={{c}_{1}}\Delta -{{\left[ s,{{\hat{H}}^{\left( cl \right)}} \right]}_{QPB}}={{c}_{1}}\left( \Delta -{{\left[ s,\Delta  \right]}_{QPB}} \right).\]It is obvious that this style of writing can be seen as a transformation of Laplace operator to a more general form defined by generalized Laplace operator, namely, $$-\Delta \to \hat{Z}= -\Delta +{{\left[ s,\Delta  \right]}_{QPB}}=-\Delta -\left( 2\nabla s\cdot \nabla +\Delta s \right)=-\Delta -\hat{Q},$$
where by simply computation ${{\left[ s,\Delta  \right]}_{QPB}}=-\left( 2\nabla s\cdot \nabla +\Delta s \right)=-\hat{Q}$.
For the sake of convenient discussions, taking the eigenvalue equation problem of generalized Laplacian
 \begin{equation}\label{a5}
   \hat{Z}=-\Delta+i{{c}_{m}}{{{\hat{w}}}^{\left( cl \right)}},
 \end{equation}
into account, where ${{c}_{m}}=-2m/\hbar \in \mathbb{R}$ is a constant number, namely, $\hat{Z}\varphi =\beta \varphi $, more precisely,  the eigenvalue equation problem of generalized
Laplacian on region $\Omega $ is
\begin{equation}\label{a13}
  -\Delta \varphi -2\nabla s\cdot \nabla \varphi -\varphi \Delta s=\beta \varphi,
\end{equation}
where $\beta =-\gamma /{{c}_{1}}$. Hence, \eqref{a13} with equipment Dirichlet
boundary conditions $\varphi \left| _{\partial \Omega } \right.=0$ naturally forms
generalized Dirichlet Laplacian \[\left\{ \begin{matrix}
   \hat{Z}\varphi =\beta \varphi ,~in ~~\Omega   \\
   \varphi \left| _{\partial \Omega }=0,~~\partial \Omega  \right.  \\
\end{matrix} \right.,\]
and \eqref{a13} equipped with the Neumann
boundary conditions $\frac{\partial \varphi }{\partial \boldsymbol {n}}\left( x \right)=0$ leads to generalized Neumann Laplacian \[\left\{ \begin{matrix}
   \hat{Z}\varphi =\beta \varphi ,~in ~~\Omega   \\
   \frac{\partial \varphi }{\partial \boldsymbol {n}}\left| _{\partial \Omega }=0,~~\partial \Omega  \right.  \\
\end{matrix} \right..\]

Actually, if we omit the constants from (2.1) to mention prominence to the key points, then (2.1) can be simply rewritten as
\begin{equation}\label{a9}
  \hat{Z}=-\Delta+i{{{\hat{w}}}^{\left( cl \right)}},
\end{equation}
for a main discussions.  As (2.2) shows, it implies a equal status for both Laplacian and the G-dynamics of type I, two operators are self-adjoint operator to construct a complex non-self-adjoint operator. Besides, Hermitian conjugate of (2.2) is deduced ${{\hat{Z}}^{\dagger }}=-\Delta -i{{\hat{w}}^{\left( cl \right)}}$, then
\begin{align}
  & -\Delta =\left( \hat{Z}+{{\hat{Z}}^{\dagger }} \right)/2, \notag\\
 & {{{\hat{w}}}^{\left( cl \right)}}=-i\left( \hat{Z}-{{\hat{Z}}^{\dagger }} \right)/2.\notag
\end{align}are both self-adjoint operators.
As a result, the connection between generalized Laplace operator and  motor operator is given by the formula ${{\hat{E}}^{\left( w \right)}}=-{{c}_{1}}\hat{Z}$. Thusly, a complete eigenvalues equation is given by
\begin{equation}\label{a6}
  -\Delta \varphi +i{{c}_{m}}{{{\hat{w}}}^{\left( cl \right)}}\varphi =\beta \varphi,
\end{equation} or the minified version of the (2.3) takes  $-\Delta \varphi +i{{{\hat{w}}}^{\left( cl \right)}}\varphi =\beta \varphi$.
Due to $\Delta$ and ${{\hat{w}}^{\left( cl \right)}}$ are self-adjoint operator, and ${{c}_{m}}$ is a real number, hence, operator $\hat{Z}$ on $\Omega $ has discrete complex eigenvalues ${{\beta }_{1}},{{\beta }_{2}},{{\beta }_{3}},\cdots $, and ${{\beta }_{k}}\left( \Omega  \right)$ means $k$-th complex eigenvalues on $\Omega $, then ${{\beta }_{k}}\left( \Omega  \right)={{\lambda }_{k}}\left( \Omega  \right)+i{{\sigma }_{k}}\left( \Omega  \right)$, where ${{\lambda }_{k}}\left( \Omega  \right),{{\sigma }_{k}}\left( \Omega  \right)$ are real eigenvalues.
In this way, we want to find the certain complex eigenvalues, to see how it represents on region $\Omega $.  In search of the certain complex eigenvalues of generalized Laplacian (2.1), we can acquire the energy spectrum ${{\gamma }_{k}}\left( \Omega  \right)=-{{c}_{1}}{{\beta }_{k}}\left( \Omega  \right)$ of motor operator (1.4) on $\Omega$ appeared in quantum mechanics as an energy operator.

Second order differential equation is a certain type of differential equation that consists of a derivative of a function of order 2 and no other higher-order derivative of the function appears in the equation.
Clearly, the eigenvalues equation \eqref{a13} of generalized Laplacian (2.1) is a standard second order differential equation with variable coefficients of the type $F\left( x,\varphi ,\nabla \varphi ,\Delta \varphi  \right)=0$ which has generalized the Laplacian $F\left( x,\varphi ,\Delta \varphi  \right)=0$, recall
\begin{equation}\label{a14}
  \Delta \varphi -i{{c}_{m}}{{\hat{w}}^{\left( cl \right)}}\varphi +\beta \varphi =\Delta \varphi +2\nabla s\cdot \nabla \varphi +\varphi \Delta s+\beta \varphi=0.
\end{equation}
Obviously, the variation of parameters works on a wider range of functions.

As we know, the second order differential equation with constant coefficients has a lot of rich researches, its research results are very mature,  for this purpose, we   firstly consider one dimensional case of \eqref{a14} with a very special condition $u=u_{0}=Const$, here geometric
potential function $s={{u}_{0}}x+{{s}_{0}}$ is a linear solution,   then a homogeneous second order differential equation with constant coefficients is of the form
\begin{equation}\label{b5}
  {{\varphi }_{xx}}+2{{u}_{0}}{{\varphi }_{x}}+\beta \varphi =0,
\end{equation}
 where ${u}_{0}$, $\beta$ are constants, the auxiliary equation or characteristic equation is
${{r}^{2}}+2{{u}_{0}}r+\beta =0$, use the quadratic equation formula
$${{r}^{1,2}}=\frac{-2{{u}_{0}}\pm \sqrt{4u_{0}^{2}-4\beta }}{2}=-{{u}_{0}}\pm \sqrt{u_{0}^{2}-\beta },$$ then two roots are respectively given by ${{r}^{1}}=-{{u}_{0}}+\sqrt{u_{0}^{2}-\beta },~~{{r}^{2}}=-{{u}_{0}}-\sqrt{u_{0}^{2}-\beta }$.

\begin{itemize}
  \item If $r^{1}$ and $r^{2}$ are real and distinct roots, $u_{0}^{2}>\beta $,  then the general solution is
$\varphi ={{A}_{1}}{{e}^{{{r}^{1}}x}}+{{B}_{1}}{{e}^{{{r}^{2}}x}}$.
  \item If $r^{1}=r^{2}=r=-{{u}_{0}}$,  that is $\beta =u_{0}^{2}$, then the general solution is
$\varphi =\left( {{A}_{1}}+{{B}_{1}}x \right){{e}^{rx}}$.
  \item
If $r^{1}=a_{1}+ib_{1}$ and $r^{2}=a_{1}-ib_{1}$  are complex roots, $u_{0}^{2}<\beta $,  then the general solution is
$\varphi ={{e}^{{{a}_{1}}x}}\left( {{A}_{1}}\sin \left( {{b}_{1}}x \right)+{{B}_{1}}\cos \left( {{b}_{1}}x \right) \right)$.
\end{itemize}
As the three solutions shown, we can see that the $u=u_{0}$ plays a key role to all different kinds of solutions, but in generally, the $u$ is not a constant but a function, this point undoubtedly enhances out difficulties to get it solution dynamically.

\subsection{Generalized Dirichlet Laplacian}
The Dirichlet eigenvalues are the fundamental modes of vibration of an idealized drum with a given shape. Accordingly, the generalized Dirichlet eigenvalues are found by solving the following problem for an unknown function $\varphi\neq0$ and complex eigenvalue $\beta$
\begin{equation}\label{a7}
  \left\{ \begin{matrix}
   -\Delta \varphi +i{{{\hat{w}}}^{\left( cl \right)}}\varphi =\beta \varphi ,\begin{matrix}
   {} & in  \\
\end{matrix}~\Omega   \\
   \varphi \left| _{\partial \Omega } \right.=0.  \\
\end{matrix} \right.
\end{equation}
where $\Delta$ is the Laplacian.
With the same reason as \eqref{a2} indicates, \eqref{a7} on $\Omega$ has discrete complex eigenvalues. The generalized Dirichlet Laplacian on $\Omega$  with Dirichlet boundary conditions is a non-self-adjoint operator, so there
exists a sequence of complex eigenvalues ${{\beta }_{1}},{{\beta }_{2}}, {{\beta}_{3}},\cdots $ and a sequence of corresponding eigenfunctions ${{\varphi }_{1}},{{\varphi }_{2}}, {{\varphi}_{3}},\cdots $. So ${{\beta }_{k}}\left( \Omega  \right)$ represents the $k$-th characteristic value of generalized Dirichlet eigenvalues problem on $\Omega$.

Notice that the Laplace operator ${\displaystyle \Delta \varphi=\operatorname {div} (\nabla \varphi)}$, then according to high dimensional Green formula, it has $${\displaystyle \int _{\varOmega }\Delta \varphi\mathrm {d} \tau=\int _{\partial \varOmega }\nabla \varphi\cdot {\boldsymbol {n}}\mathrm {d}S(x).}$$ In general, $${\displaystyle \int _{\varOmega }\phi\Delta \varphi\mathrm {d} \tau=-\int _{\varOmega }\nabla \varphi\cdot\nabla \phi\mathrm {d} \tau+\int _{\partial \varOmega }{\dfrac {\partial \phi}{\partial {\boldsymbol {n}}}}\varphi\mathrm {d}S(x).}$$where ${\displaystyle {\boldsymbol {n}}=(n_{1},n_{2},\cdots ,n_{N})}$ is the normal vector of ${\displaystyle \partial \varOmega }$.

\begin{theorem}\label{t1}
  The G-dynamics of type I in the generalized Dirichlet Laplacian case, it has
$$\int\limits_{\Omega }{\left( {{{\hat{w}}}^{\left( cl \right)}}\varphi  \right)\varphi d\tau} =0,$$where $\varphi \left| _{\partial \Omega } \right.=0$.
\begin{proof}
  Let $\varphi$ be an eigenfunction.
   Based on ${{\hat{w}}^{\left( cl \right)}}\varphi ={{b}_{c}}\left( 2\nabla s\cdot \nabla \varphi +\varphi \Delta s \right)$, then the integral calculation is given by
\begin{align}
 \int\limits_{\Omega }{\left( {{{\hat{w}}}^{\left( cl \right)}}\varphi  \right)\varphi d\tau} &={{b}_{c}}\int\limits_{\Omega }{\left( 2\nabla s\cdot \nabla \varphi +\varphi \Delta s \right)\varphi d\tau} \notag\\
 & =2{{b}_{c}}\int\limits_{\Omega }{\left( \nabla s\cdot \nabla \varphi  \right)\varphi d\tau}-{b}_{c}\int\limits_{\Omega }{\left( \nabla s\cdot \nabla {{\varphi }^{2}} \right)d\tau}+{b}_{c}\int\limits_{\partial \Omega }{s\frac{\partial {{\varphi }^{2}}}{\partial  \boldsymbol {n}}dS\left( x \right)} \notag\\
 & =2{{b}_{c}}\int\limits_{\Omega }{\left( \nabla s\cdot \nabla \varphi  \right)\varphi d\tau-2}{{b}_{c}}\int\limits_{\Omega }{\left( \nabla s\cdot \nabla \varphi  \right)\varphi d\tau} \notag\\
 & =0 ,\notag
\end{align}
where the divergence theorem is used, we have
\begin{align}
 \int\limits_{\Omega }{\left( {{\varphi }^{2}}\Delta s \right)d\tau} & =-\int\limits_{\Omega }{\left( \nabla s\cdot \nabla {{\varphi }^{2}} \right)d\tau}+\int\limits_{\partial \Omega }{s\frac{\partial {{\varphi }^{2}}}{\partial  \boldsymbol {n}}dS\left( x \right)}  \notag\\
 & =-\int\limits_{\Omega }{\left( \nabla s\cdot \nabla {{\varphi }^{2}} \right)d\tau} , \notag
\end{align}
where $\varphi$ is a function that vanishes on the boundary of $\Omega$. The proof is done.
\end{proof}
\end{theorem}
In fact,  for the eigenvalue problem (2.4) above, it can be derived
\begin{align}\label{a10}
 -\int\limits_{\Omega }{\left( \Delta \varphi -i{{{\hat{w}}}^{\left( cl \right)}}\varphi  \right)\varphi d\tau} &=-\int\limits_{\Omega }{\left( \Delta \varphi  \right)\varphi d\tau}+i\int\limits_{\Omega }{\left( {{{\hat{w}}}^{\left( cl \right)}}\varphi  \right)\varphi d\tau} \notag\\
 & =\int\limits_{\Omega }{{{\left| \nabla \varphi  \right|}^{2}}d\tau}-\int\limits_{\partial \Omega }{\varphi \frac{\partial \varphi }{\partial  \boldsymbol {n}}dS\left( x \right)}\\
 & \begin{matrix}
   {} & {} & {} & {}  \\
\end{matrix}+2i{{b}_{c}}\int\limits_{\Omega }{\left( \nabla s\cdot \nabla \varphi  \right)\varphi d\tau}+i{{b}_{c}}\int\limits_{\Omega }{\left( {{\varphi }^{2}}\Delta s \right)d\tau} \notag\\
 & =\int\limits_{\Omega }{{{\left| \nabla \varphi  \right|}^{2}}d\tau}-\int\limits_{\partial \Omega }{\left( 1-2i{{b}_{c}}s \right)\varphi \frac{\partial \varphi }{\partial  \boldsymbol {n}}dS\left( x \right)} \notag\\
 & =\int\limits_{\Omega }{{{\left| \nabla \varphi  \right|}^{2}}d\tau}\ge 0, \notag
\end{align}
where theorem \ref{t1} is used during the derivation, and boundary condition  $$\int\limits_{\partial \Omega }{\left( 1-2i{{b}_{c}}s \right)\varphi \frac{\partial \varphi }{\partial  \boldsymbol {n}}dS\left( x \right)}=0,$$ holds.
Using \eqref{a10}, it is proved that real parts ${{\lambda }_{k}}\left( \Omega  \right)$ of all eigenvalues are positive in the generalized Dirichlet Laplacian case (2.4).

\subsection{Generalized Neumann Laplacian}

Consider a bounded domain $\Omega \subset {{\mathbb{R}}^{n}}$ with a piecewise smooth boundary $\partial \Omega$. $\mu$ is a Neumann eigenvalue on $\Omega$ if there exists a Neumann eigenfunction $\varphi \in {{C}^{2}}\left( \Omega  \right)\cap {{C}^{0}}\left( \overline{\Omega } \right)$ such that
\begin{equation}\label{a8}
  \left\{ \begin{matrix}
   -\Delta \varphi \left( x \right)+i{{{\hat{w}}}^{\left( cl \right)}}\varphi\left( x \right)=\mu \left( x \right)\varphi \left( x \right),~x\in \Omega   \\
   \frac{\partial \varphi }{\partial \boldsymbol {n}}\left( x \right)=0,~x\in \partial \Omega .  \\
\end{matrix} \right.
\end{equation}holds,
where $\frac{\partial \varphi }{\partial \boldsymbol {n}}\left( x \right)$ is the normal derivative of $\varphi$.   The generalized Neumann Laplacian has a discrete complex spectrum:
$\mu_{1}(\Omega),\mu_{2}(\Omega),\cdots$, ~
${{\mu }_{k}}\rightarrow\infty$ as $k\rightarrow\infty $.

Note that the formula \[\int\limits_{\partial \Omega }{\varphi \frac{\partial \varphi }{\partial  \boldsymbol {n}}dS\left( x \right)}=\int\limits_{\Omega }{\left( \varphi \Delta \varphi  \right)d\tau}+\int\limits_{\Omega }{{{\left| \nabla \varphi  \right|}^{2}}d\tau},\] is useful to the boundary condition such as both Dirichlet boundary conditions $\varphi \left| _{\partial \Omega } \right.=0$ and Neumann boundary conditions $\frac{\partial \varphi }{\partial \boldsymbol {n}}\left( x \right)=0,~x\in \partial \Omega$, with the help of the boundary terms that simultaneously appear in this integal term
$\int\limits_{\partial \Omega }{\varphi \frac{\partial \varphi }{\partial \boldsymbol {n}}dS\left( x \right)}$.

\subsection{One-dimensional case of eigenvalues equation}

\begin{theorem}
 The one-dimensional G-dynamics of type I satisfies
 $$\int\limits_{\Omega }{\left( {{{\hat{w}}}^{\left( cl \right)}}\varphi  \right)\varphi dx}=0.$$

\begin{proof}
The one-dimensional G-dynamics of type I leads to  ${{\hat{w}}^{\left( cl \right)}}\varphi ={{b}_{c}}\left( 2u{{\varphi }_{x}}+\varphi {{u}_{x}} \right)$, then by a direct computation, it gets
  \begin{align}
  \int\limits_{\Omega }{\left( {{{\hat{w}}}^{\left( cl \right)}}\varphi  \right)\varphi dx}&={{b}_{c}}\int\limits_{\Omega }{\left( 2u{{\varphi }_{x}}+\varphi {{u}_{x}} \right)\varphi dx} \notag\\
 & =2{{b}_{c}}\int\limits_{\Omega }{u{{\varphi }_{x}}\varphi dx}+{{b}_{c}}\int\limits_{\Omega }{{{u}_{x}}{{\varphi }^{2}}dx} \notag\\
 & =2{{b}_{c}}\int\limits_{\Omega }{ud\left( \frac{1}{2}{{\varphi }^{2}} \right)}+{{b}_{c}}\int\limits_{\Omega }{{{\varphi }^{2}}du} \notag\\
 & ={{b}_{c}}u{{\varphi }^{2}}\left| _{\partial\Omega } \right.-{{b}_{c}}\int\limits_{\Omega }{{{\varphi }^{2}}du}+{{b}_{c}}\int\limits_{\Omega }{{{\varphi }^{2}}du} \notag\\
 & ={{b}_{c}}u{{\varphi }^{2}}\left| _{\partial\Omega } \right.\notag\\
 &=0,\notag
\end{align}
where we have used integration by parts, and
threw away the boundary term for the usual reason: if $\varphi$ is square
integrable, they must go to zero at $\pm\infty$. The proof is complete.

\end{proof}
\end{theorem}
Generally, a second order differential equation with variable coefficients is written as
\begin{equation}\label{c2}
  {{\varphi }_{xx}}+ r_{1}(x){{\varphi}_{x}} + h_{1}(x){{\varphi }} = f(x)
\end{equation}
where $r_{1}(x), h_{1}(x)$, and $f(x)$ are functions of $x$. it can be solved this differential equation by using the auxiliary equation and different methods such as the method of undetermined coefficients and variation of parameters.
The one-dimensional motor operator is
 $${{{\hat{E}}}^{\left( w \right)}}={{c}_{1}}{d}^{2}/d{{x}^{2}}-i\hbar {{\hat{w}}^{\left( cl \right)}}={{c}_{1}}\left( {{d}^{2}}/d{{x}^{2}}+\hat{Q} \right).$$Then it leads to $-{{\hat{E}}^{\left( w \right)}}/{{c}_{1}}=-{{d}^{2}}/d{{x}^{2}}-\hat{Q}$ that fits a second order differential equation with variable coefficients when acts onto a function.
  The general form of a second-order differential equation is
  \begin{equation}\label{c4}
    F\left( x,\varphi ,{{\varphi}_{x}},{{\varphi}_{xx}} \right)=0.
  \end{equation}
The one-dimensional generalized Laplacian can be formally rewritten as
\begin{equation}\label{c3}
 {{\varphi }_{xx}}+2u{{\varphi }_{x}}+{{u}_{x}}\varphi +\beta \varphi =0.
\end{equation}
 In order to show the second order differential equation with variable coefficients \eqref{c2} with the one-dimensional generalized Laplacian \eqref{c3}, putting them together for a comparison,
 $$\left\{ \begin{matrix}
   {{\varphi }_{xx}}+{{r}_{1}}(x){{\varphi }_{x}}+{{h}_{1}}(x)\varphi =f(x)  \\
    {{\varphi }_{xx}}-i{{c}_{m}}{{\hat{w}}^{\left( cl \right)}}\varphi={{\varphi }_{xx}}+2u(x){{\varphi }_{x}}+{{u}_{x}}(x)\varphi =-\beta \varphi \left( x \right)  \\
\end{matrix} \right.$$
It can be treated as following equalities:
\begin{align}
  & {{r}_{1}}(x)=2u(x), \notag\\
 & {{h}_{1}}(x)={{u}_{x}}(x), \notag\\
 & f(x)=-\beta \varphi \left( x \right). \notag
\end{align}As shown, the variable coefficients are the function $u$ and its first derivative in terms of $x$.  Both \eqref{c2} and \eqref{c3} can be summarized as the form of equation \eqref{c4}. That's why the Laplacian needs to be extended to the generalized Laplacian \eqref{c3} which can explain more wider system with some boundaries, in particular, the generalized Laplacian \eqref{c3} directly links to the energy spectrum in quantum mechanics such as the energy spectrum of the motor operator.
It helps to study the following one-dimensional eigenvalues equation of generalized Dirichlet Laplacian (2.4) in a certain form that is
\begin{equation}\label{a11}
  \left\{ \begin{matrix}
   -{{d}^{2}}\psi /d{{x}^{2}}+i{{{\hat{w}}}^{\left( cl \right)}}\psi =\beta \psi ,\begin{matrix}
   {} & in  \\
\end{matrix}~~\Omega   \\
   \psi \left| _{\partial \Omega } \right.=0,  \\
\end{matrix} \right.
\end{equation}
In the beginning, we start with the one-dimensional eigenvalue equation
of the G-dynamics of the type I,
consider the eigenvalue equation ${{\hat{w}}^{\left( cl \right)}}{{\psi }_{n}}=w_{n}^{\left( q \right)}{{\psi }_{n}}$, by solving the eigenvalues equation,
\begin{equation}\label{b1}
  \frac{2u}{{{\psi }_{n}}}\frac{d{{\psi }_{n}}}{dx}+{{u}_{x}}=w_{n}^{\left( q \right)}/{{b}_{c}},
\end{equation} By direct computation, the general solution of the $n$-stationary wave function of the system can be given by
\begin{equation}\label{a12}
  {{\psi }_{n}}={{C}_{0}}{{u}^{-1/2}}{{e}^{iw_{n}^{\left( q \right)}a_{c}^{-1}\int{\frac{dx}{u}}}},~~u>0,
\end{equation}
where $w_{n}^{\left( q \right)}$ are the eigenvalue equations in terms of the eigenfunctions \eqref{a12} ${{\psi }_{n}}$ given above, where ${{a}_{c}}=\hbar /m$. the eigenfunction can be simply rewritten as
${{\psi }_{n}}={{C}_{0}}{{u}^{-1/2}}{{e}^{iw_{n}^{\left( q \right)}t}}$, where ${C}_{0}$ is a constant, and we have used \[t=a_{c}^{-1}\int{\frac{dx}{u}}={{c}_{0}}\int{\frac{dx}{\dot{x}}}={{c}_{0}}\int{dt},\]
to express the period related to the frequency spectrum $w_{n}^{\left( q \right)}$, where the one-dimensional line element $dx=vdt=\dot{x}dt$ is applied to above derivation, and the one-dimensional velocity $v=\dot{x}=dx/dt$,  while ${{c}_ {0}}=a_ {c} ^{-1}v/u$ is a non-zero real constant ratio, generally $\pm 1$. Correspondingly, \[\left\{ \begin{matrix}
   {{{\hat{w}}}^{\left( cl \right)}}{{\psi }_{n}}=w_{n}^{\left( q \right)}{{\psi }_{n}},  \\
   {{\psi }_{n}}={{C}_{0}}{{u}^{-1/2}}{{e}^{iw_{n}^{\left( q \right)}t}},  \\
   w_{n}^{\left( q \right)}\in \mathbb{R},u>0.  \\
\end{matrix} \right.\]
by studying the Laplacian, it reveals the G-dynamics of the type I has a discrete spectrum of infinitely many positive eigenvalues:
 $0<w_{1}^{\left( q \right)}< w_{2}^{\left( q \right)}<w_{3}^{\left( q \right)}<\cdots $, and a sequence of corresponding eigenfunctions ${{\psi }_{1}},{{\psi}_{2}}, {{\psi}_{3}},\cdots $.

As the eigenfunction given by the G-dynamics of the type I certainly shows,
for this accurate expression \eqref{a12}, we can estimate the eigenvalues of the \eqref{a11} in taking this specific eigenfunction into consideration.
Consider
\begin{equation}\label{b9}
  -{{d}^{2}}\psi_{n} /d{{x}^{2}}+i{{c}_{m}}{{{\hat{w}}}^{\left( cl \right)}}\psi_{n} =\beta_{n} \psi_{n}.
\end{equation}
Firstly, to evaluate ${{d}^{2}}\psi_{n} /d{{x}^{2}}$, due to \eqref{b1}, it has  $d\psi_{n} /dx=\eta_{n} \psi_{n}$, where $\eta_{n} =ia_{c}^{-1}\frac{w_{n}^{\left( q \right)}}{u}-\frac{1}{2 u}{{u}_{x}}$, therefore, ${{d}^{2}}\psi_{n} /d{{x}^{2}}=\left( {{\eta_{n} }^{2}}+d\eta_{n} /dx \right)\psi_{n} $, through calculation and sorting, there has
$$-{{d}^{2}}\psi_{n} /d{{x}^{2}}=\left( a_{c}^{-2}\frac{{w_{n}^{\left( q \right)2}}}{{{u}^{2}}}+2ia_{c}^{-1}\frac{w_{n}^{\left( q \right)}{{u}_{x}}}{{{u}^{2}}}-\frac{3}{4}\frac{{{u}^{2}}_{x}}{{{u}^{2}}}+\frac{1}{2}\frac{{{u}_{xx}}}{u} \right)\psi_{n}. $$
Therefore, the complex eigenvalues are given by
\begin{equation}\label{c6}
  \beta_{n} =\frac{1}{{{u}^{2}}}\left( a_{c}^{-2}{w_{n}^{\left( q \right)2}}-\frac{3}{4}{{u}^{2}}_{x}+\frac{1}{2}u{{u}_{xx}} \right)+i\left( 2a_{c}^{-1}{w_{n}^{\left( q \right)}}\frac{{{u}_{x}}}{{{u}^{2}}}-\frac{2m}{\hbar }{w_{n}^{\left( q \right)}} \right).
\end{equation}
corresponding to the eigenfunction \eqref{a12},
where the real part and imaginary part of the
eigenvalues of the complex eigenvalues \eqref{c6} are separately given by   $$\operatorname{Re}{{\beta }_{n}}={{\lambda }_{n}}\left( \Omega  \right)=\frac{1}{{{u}^{2}}}\left( a_{c}^{-2}{w_{n}^{\left( q \right)2}}-\frac{3}{4}{{u}^{2}}_{x}+\frac{1}{2}u{{u}_{xx}} \right),$$~~$$\operatorname{Im}{{\beta}_{n}}={{\sigma }_{n}}\left( \Omega  \right)=2a_{c}^{-1}{w_{n}^{\left( q \right)}}\frac{{{u}_{x}}}{{{u}^{2}}}-\frac{2m}{\hbar }{w_{n}^{\left( q \right)}}. $$

Obviously, the complex eigenvalues $\beta_{n}$ is a little complex, owing to the ${w_{n}^{\left( q \right)}}$ and function $u$ with its derivative. Given a special case of positive function $u$,
supposed ${{u}_{x}}=0$, that is, $u={{u}_{0}}>0$ is a constant, then the complex eigenvalues are written in a simple form
\[{{\beta }_{0n}}={{\left( \frac{w_{n}^{\left( q \right)}}{{{a}_{c}}{{u}_{0}}} \right)}^{2}}+ic_{m}w_{n}^{\left( q \right)},\]corresponding to the eigenfunction \eqref{a12}, where  ${{\lambda }_{0n}}\left( \Omega  \right)={{\left( \frac{w_{n}^{\left( q \right)}}{{{a}_{c}}{{u}_{0}}} \right)}^{2}},~~{{\sigma }_{n}}\left( \Omega  \right)=c_{m}w_{n}^{\left( q \right)}$. Accordingly, it obtains the energy spectrum ${{\gamma }_{0n}}\left( \Omega  \right)=-{{c}_{1}}{{\beta }_{0n}}\left( \Omega  \right)$ of motor operator (1.4) on $\Omega$. More precisely,
\begin{equation}\label{c1}
  {{\gamma }_{0n}}=-{{c}_{1}}{{\beta }_{0n}}=-{{c}_{1}}{{\left( \frac{w_{n}^{\left( q \right)}}{{{a}_{c}}{{u}_{0}}} \right)}^{2}}-i\hbar w_{n}^{\left( q \right)}=mv_{0n}^{2}/2-iE_{n}^{\left( q \right)},
\end{equation}
where ${{v}_{0n}}=w_{n}^{\left( q \right)}/{{u}_{0}},~~E_{n}^{\left( q \right)}=\hbar w_{n}^{\left( q \right)}$. 
Therefore, it gets the real part and imaginary part of the eigenvalues \eqref{c1} expressed as
\begin{align}
  & \operatorname{Re}{{\gamma }_{0n}}=mv_{0n}^{2}/2, \notag\\
 & \operatorname{Im}{{\gamma }_{0n}}=-E_{n}^{\left( q \right)}. \notag
\end{align}
As it shows, the complex eigenvalues ${{\gamma }_{0n}}$ are linked to the real eigenvalues $w_{n}^{\left( q \right)}$ deduced by the G-dynamics of the type I.
It is clear to see that such special case gives us another vision to see how exactly the eigenvalues and its corresponding eigenfunctions might be. So in this special case, we can get the specific eigenvalues with the concrete eigenfunctions \eqref{a12} for the  motor operator totally expressed as follows:
$$\left\{ \begin{matrix}
   {{{\hat{E}}}^{\left( w \right)}}{{\psi }_{n}}={{c}_{1}}{{d}^{\text{2}}}{{\psi }_{n}}/d{{x}^{\text{2}}}-i\hbar {{{\hat{w}}}^{\left( cl \right)}}{{\psi }_{n}}={{\gamma }_{0n}}{{\psi }_{n}} , \\
   {{\psi }_{n}}={{C}_{0}}{{u}^{-1/2}}{{e}^{iw_{n}^{\left( q \right)}t}} , \\
  {\gamma }_{0n}= {{\gamma }_{n}}\left| _{u={{u}_{0}}} \right.=mv_{0n}^{2}/2-iE_{n}^{\left( q \right)} . \\
\end{matrix} \right.$$
By the way, the eigenvalues for classical kinetic operator with eigenfunctions \eqref{a12} is written by
${{c}_{1}}{{d}^{\text{2}}}{{\psi }_{n}}/d{{x}^{\text{2}}}\left| _{{{u}_{x}}=0} \right.=mv_{0n}^{2}{{\psi }_{n}}/2$. As we can see, the eigenvalues for classical kinetic operator are in a classical expression $mv_{0n}^{2}/2$, that means classical kinetic energy $T_{0n}^{\left( cl \right)}=\frac{m}{2}v_{0n}^{2}$, namely, ${{c}_{1}}{{d}^{\text{2}}}{{\psi }_{n}}/d{{x}^{\text{2}}}\left| _{{{u}_{x}}=0} \right.=T_{0n}^{\left( cl \right)}\psi_{n}$, it indicates that classical kinetic energy has been discretized under the corresponding eigenfunctions \eqref{a12}.

In general case of \eqref{c1}, that is to say, the condition $u_{x}\neq0$ holds for the energy spectrum of the motor operator,  taking \eqref{c6} into consideration, then it yields the general result
 \begin{align}\label{c7}
{{\gamma }_{n}}  & =-{{c}_{1}}{{\beta }_{n}} =\frac{m}{2}{{v}_{n}}^{2}+\frac{3}{4}{{c}_{1}}\frac{u_{x}^{2}}{{{u}^{2}}}-\frac{{{c}_{1}}}{2}\frac{{{u}_{xx}}}{u}
-iE_{n}^{\left( q \right)}\left( 1-\frac{{{u}_{x}}}{{{u}^{2}}} \right),
\end{align}which is corresponding to the eigenfunction \eqref{a12},
where ${{v}_{n}}=w_{n}^{\left( q \right)}/{{u}}$ has been used.  As a result, it gets the real part and imaginary part of the eigenvalues of the motor operator \eqref{c7}  given by
\begin{align}
  & \operatorname{Re}{{\gamma }_{n}}=T_{n}^{\left( cl \right)}+\frac{3}{4}{{c}_{1}}\frac{u_{x}^{2}}{{{u}^{2}}}
  -\frac{{{c}_{1}}}{2}\frac{{{u}_{xx}}}{u}, \notag\\
 & \operatorname{Im}{{\gamma }_{n}}=-E_{n}^{\left( q \right)}\left( 1-\frac{{{u}_{x}}}{{{u}^{2}}} \right)=-E_{n}^{\left( q \right)}+E_{n}^{\left( q \right)}\frac{{{u}_{x}}}{{{u}^{2}}}, \notag
\end{align}where $T_{n}^{\left( cl \right)}=\frac{m}{2}v_{n}^{2}$, therefore, if $u_{x}=0$, then general situation \eqref{c7} is degenerated to the simple case \eqref{c1}.

\section{Type II and type III for the generalized Laplacian}
Above all, it's dealt with the case above associated with the G-dynamics of the type I which the generalized Laplacian equipped with,  but there are three different kinds of G-dynamics, the G-dynamics of the type I in the generalized Laplacian can be replaced by the one-dimensional G-dynamics of type II, namely,
\begin{equation}\label{b7}
  {{\hat{Z}}^{\left( II \right)}}=-\Delta +i{{c}_{m}}{{\hat{w}}^{\left( ri \right)}}.
\end{equation}
Notice that this way of writing characters is aimed to distinguish generalized Laplacian with G-dynamics of the type I.  More precisely, it contains the G-dynamics of the type I and type III simultaneously, hence, generalized Laplacian becomes
\begin{equation}\label{b8}
  {{\hat{Z}}^{\left( II \right)}}=-\Delta +i{{c}_{m}}{{\hat{w}}^{\left( ri \right)}}=-\Delta +i{{c}_{m}}{{\hat{w}}^{\left( cl \right)}}+i{{c}_{m}}{{w}^{\left( s \right)}}.
\end{equation}
As a consequence, it has ${{\hat{Z}}^{\left( II \right)}}=\hat{Z}+i{{c}_{m}}{{w}^{\left( s \right)}}$. It needs to be emphasized  that ${{\hat{w}}^{\left( ri \right)}}$ is a non-Hermitian operator, more accurate, where $\Delta, ~{{\hat{w}}^{\left( cl \right)}}$ are Hermitian operators while
${{w}^{\left( s \right)}}$ is an anti-Hermitian operator,  that means $i{{w}^{\left( s \right)}}$ is a Hermitian operator.
This can be treated as the more general form of generalized Laplacian, taking \eqref{a3} into the formula of the G-dynamics leads to the result
\begin{align}
i\hbar {{\hat{w}}^{\left( ri \right)}}&= {{\left[ s,{{\hat{H}}^{\left( ri \right)}} \right]}_{QPB}}={{\left[ s,{{\hat{H}}^{\left( cl \right)}}-i\hbar {{\hat{w}}^{\left( cl \right)}} \right]}_{QPB}}\notag\\
&={{\left[ s,{{\hat{H}}^{\left( cl \right)}} \right]}_{QPB}}-i\hbar {{\left[ s,{{\hat{w}}^{\left( cl \right)}} \right]}_{QPB}} \notag\\
 & =-{{c}_{1}}\left( 2\nabla s\cdot \nabla +\Delta s+2{{\left| \nabla s \right|}^{2}} \right). \notag
\end{align}In support of the G-dynamics of the type II and the Laplacian, it results in the G-operator as \eqref{b6} expressed that is given by
\begin{align}
 {{\hat{H}}^{\left( gr \right)}} -V&={{c}_{1}}\Delta -{{\left[ s,{{\hat{H}}^{\left( ri \right)}} \right]}_{QPB}}={{c}_{1}}\Delta -i\hbar {{\hat{w}}^{\left( ri \right)}}\notag\\
 & ={{c}_{1}}\left( \Delta +2\nabla s\cdot \nabla +\Delta s+{{\left| \nabla s \right|}^{2}} \right)+{{c}_{1}}{{\left| \nabla s \right|}^{2}} \notag\\
 & ={{\hat{T}}^{\left( ri \right)}}+{{c}_{1}}{{\left| \nabla s \right|}^{2}}. \notag
\end{align}
where the geometrinetic energy operator here takes the form \[{{\hat{T}}^{\left( ri \right)}}={{c}_{1}}\left( \Delta +2\nabla s\cdot \nabla +\Delta s+{{\left| \nabla s \right|}^{2}} \right).\]
For this case, it's clear to obtain generalized Laplacian related to the G-dynamics of the type II, shown by the transformation
\[-\Delta \to {{\hat{Z}}^{\left( II \right)}}=-\Delta -\left( 2\nabla s\cdot \nabla +\Delta s+2\nabla s\cdot \nabla s \right).\]
As a result, the eigenvalues equation of generalized Laplacian related to the G-dynamics of the type II follows ${{\hat{Z}}^{\left( II \right)}}\varphi ={{\beta }^{\left( II \right)}}\varphi$, it shows in details
\begin{equation}\label{b3}
  \hat{Z}\varphi +i{{c}_{m}}{{w}^{\left( s \right)}}\varphi =-\Delta \varphi +i{{c}_{m}}{{\hat{w}}^{\left( cl \right)}}\varphi +i{{c}_{m}}{{w}^{\left( s \right)}}\varphi ={{\beta }^{\left( II \right)}}\varphi.
\end{equation}
Once the eigenvalues ${{\beta }^{\left( II \right)}}$ verify, then the energy spectrum of
the G-operator can be given by ${{\hat{H}}^{\left( gr \right)}}\varphi ={{E}^{\left( gr \right)}}\varphi $, their relation is 
\begin{equation}\label{d2}
  {{\hat{H}}^{\left( gr \right)}}=-{{c}_{1}}{{\hat{Z}}^{\left( II \right)}}+V,
\end{equation}
and the relation of the complex eigenvalues is ${{E}^{\left( gr \right)}}=-{{c}_{1}}{{\beta }^{\left( II \right)}}+V$ in which $V$ is the potential energy, on the contrary, the eigenvalues \eqref{b3} of generalized Laplacian related to the type II are shown by \[{{\beta }^{\left( II \right)}}=-\left( {{E}^{\left( gr \right)}}-V \right)/{{c}_{1}}.\]
The equation \eqref{b3} can be rewritten as a standard second order differential equation with variable coefficients of the type $F\left( x,\varphi ,\nabla \varphi ,\Delta \varphi  \right)=0$, that is shown by \[\Delta \varphi -i{{c}_{m}}{{\hat{w}}^{\left( cl \right)}}\varphi -i{{c}_{m}}{{w}^{\left( s \right)}}\varphi +{{\beta }^{\left( II \right)}}\varphi =0,\]or in the form
\begin{equation}\label{b4}
  \Delta \varphi +2\nabla s\cdot \nabla \varphi +\varphi \Delta s+2{{\left| \nabla s \right|}^{2}}\varphi+{{\beta }^{\left( II \right)}}\varphi=0.
\end{equation}
On the foudation of this second order differential equation, then when it equips with the different boundary conditions, it gives varied cases, such as generalized Dirichlet Laplacian and generalized Neumann Laplacian, etc.
\[\left\{ \begin{matrix}
   {{{\hat{Z}}}^{\left( II \right)}}\varphi =\hat{Z}\varphi +i{{c}_{m}}{{w}^{\left( s \right)}}\varphi ={{\beta }^{\left( II \right)}}\varphi ,~~in~~\Omega ,  \\
   \varphi \left| _{\partial \Omega }=0,\begin{matrix}
   {} & {}  \\
\end{matrix}/\frac{\partial \varphi }{\partial \boldsymbol {n}}\left( x \right) =0,~~\partial \Omega  \right.  \\
\end{matrix} \right..\]
Transparently, we can see that this adds the G-dynamics of the type III to the generalized Laplacian \eqref{a5} with boundary conditions, due to the G-dynamics of the type II contains the G-dynamics of the type I and type III.

\subsection{One-dimensional eigenvalues equation}
The corresponding one-dimensional eigenvalues equation of second order differential equation \eqref{b4} is given by
\begin{equation}\label{c5}
  {{\varphi }_{xx}}+2u{{\varphi }_{x}}+\varphi {{u}_{x}}+2{{u}^{2}}\varphi +{{\beta }^{\left( II \right)}}\varphi =0,
\end{equation}
it fits the universal form $F\left( x,\varphi ,{{\varphi}_{x}},{{\varphi}_{xx}} \right)=0$. Let second order differential equation \eqref{c5} be in comparison with \eqref{c2}, it gets
\begin{align}
  & {{r}_{1}}(x)=2u(x), \notag\\
 & {{h}_{1}}(x)={{u}_{x}}(x)+2{{u}^{2}}(x), \notag\\
 & f(x)=-{{\beta }^{\left( II \right)}}\varphi \left( x \right). \notag
\end{align}Here the variable coefficients are related to function $u(x)$ and its derivative ${{u}_{x}}(x)$. As previously \eqref{b5} shows, consider a special case ${{u}_{x}}=0$, then $u={{u}_{0}}>0$, therefore, under this special condition, it has a special second order differential equation expressed as \[{{\varphi }_{xx}}+2{{u}_{0}}{{\varphi }_{x}}+2u_{0}^{2}\varphi +{{\beta }^{\left( II \right)}}\varphi =0.\]
The characteristic equation is
${{r}^{2}}+2{{u}_{0}}r+2u_{0}^{2}+{{\beta }^{\left( II \right)}} =0$, using the quadratic equation formula
$${{r}^{1,2}}=-{{u}_{0}}\pm \sqrt{-u_{0}^{2}-{{\beta }^{\left( II \right)}} },$$ it needs to be stressed that if ${{r}^{1}}={{r}^{2}}=-{{u}_{0}}$, then ${{\beta }^{\left( II \right)}}=-u_{0}^{2}<0$ is a negative value which is the opposite result of \eqref{b5}.

Let's turn to the one-dimensional eigenvalues equation of \eqref{b7} corresponding to the eigenfunctions \eqref{a12} ${{\psi }_{n}}$, that is, ${{\hat{Z}}^{\left( II \right)}}{{\psi }_{n}}=\beta _{n}^{\left( II \right)}{{\psi }_{n}}$, more precisely,
\[-\Delta {{\psi }_{n}}+i{{c}_{m}}{{\hat{w}}^{\left( ri \right)}}{{\psi }_{n}}=-\Delta {{\psi }_{n}}+i{{c}_{m}}{{\hat{w}}^{\left( cl \right)}}{{\psi }_{n}}+i{{c}_{m}}{{w}^{\left( s \right)}}{{\psi }_{n}}=\beta _{n}^{\left( II \right)}{{\psi }_{n}}.\]
As a result of the one-dimensional eigenvalues equation
of the G-dynamics of the type I, and using \eqref{b9}, it leads to the outcome
\[{{\beta }_{n}}{{\psi }_{n}}+i{{c}_{m}}{{w}^{\left( s \right)}}{{\psi }_{n}}=\beta _{n}^{\left( II \right)}{{\psi }_{n}}.\]Thusly, the eigenvalues $\beta _{n}^{\left( II \right)}$ are given by \[\beta _{n}^{\left( II \right)}={{\beta }_{n}}+i{{c}_{m}}{{w}^{\left( s \right)}}={{\beta }_{n}}-2{{u}^{2}},\]where $i{{c}_{m}}{{w}^{\left( s \right)}}=-2{{u}^{2}}$.
Therefore, the complex eigenvalues of eigenvalues equation of \eqref{b7} are given by
\begin{align}\label{d1}
  & \beta _{n}^{\left( II \right)}=\frac{1}{{{u}^{2}}}\left( a_{c}^{-2}w_{n}^{\left( q \right)2}-\frac{3}{4}{{u}^{2}}_{x}+\frac{1}{2}u{{u}_{xx}} \right)-2{{u}^{2}}\notag \\
 & \begin{matrix}
   \begin{matrix}
   {} & {}  \\
\end{matrix} & {} & {} & {} & {} & {} & {} & {} & {} & {}  \\
\end{matrix}+i\left( 2a_{c}^{-1}w_{n}^{\left( q \right)}\frac{{{u}_{x}}}{{{u}^{2}}}-\frac{2m}{\hbar }w_{n}^{\left( q \right)} \right), 
\end{align}
where the real part and imaginary part of the
eigenvalues of the complex eigenvalues \eqref{d1} are respectively expressed as   
\begin{align}
  & \operatorname{Re}\beta _{n}^{\left( II \right)}=\frac{1}{{{u}^{2}}}\left( a_{c}^{-2}{w_{n}^{\left( q \right)2}}-\frac{3}{4}{{u}^{2}}_{x}+\frac{1}{2}u{{u}_{xx}} \right)-2{{u}^{2}}, \notag\\
 & \operatorname{Im}\beta _{n}^{\left( II \right)}=  2a_{c}^{-1}{w_{n}^{\left( q \right)}}\left( \frac{{{u}_{x}}}{{{u}^{2}}}-1\right). \notag
\end{align}
Analogously, using the special condition ${{u}_{x}}=0$, then, $u={{u}_{0}}>0$ is a constant, it greatly simplifies the complex eigenvalues of \eqref{b7},
\[\beta _{0n}^{\left( II \right)}={{\left( \frac{{{v}_{0n}}}{{{a}_{c}}} \right)}^{2}}-2u_{0}^{2}+ic_{m}w_{n}^{\left( q \right)},\]
where the real part ${{\lambda }_{0n}}\left( \Omega  \right)={{\left( \frac{{{v}_{0n}}}{{{a}_{c}}} \right)}^{2}}-2u_{0}^{2}$, and the imaginary part ${{\sigma }_{n}}\left( \Omega  \right)=c_{m}w_{n}^{\left( q \right)}$. In such a particular case, the eigenvalues $\beta _{0n}^{\left( II \right)}$ are verified once,  the energy spectrum of
the G-operator then follows ${{\hat{H}}^{\left( gr \right)}}{{\psi }_{n}}=E_{0n}^{\left( gr \right)}{{\psi }_{n}} $, therefore, the complex eigenvalues transform to the energy spectrum as follows:
\begin{align}\label{d3}
 E_{0n}^{\left( gr \right)}=-{{c}_{1}}\beta _{0n}^{\left( II \right)}+V&=-{{c}_{1}}{{\left( \frac{{{v}_{0n}}}{{{a}_{c}}} \right)}^{2}}+2{{c}_{1}}u_{0}^{2}+V-i\hbar w_{n}^{\left( q \right)}\notag\\
 & =mv_{0n}^{2}/2+V-{{E}^{\left( s \right)}}\left( {{u}_{0}} \right)-iE_{n}^{\left( q \right)}, \\
 & =E_{0n}^{\left( cl \right)}-{{E}^{\left( s \right)}}\left( {{u}_{0}} \right)-iE_{n}^{\left( q \right)}, \notag
\end{align}where ${{E}^{\left( s \right)}}\left( {{u}_{0}} \right)=2{{c}_{1}}u_{0}^{2}$ and the eigenvalues $$E_{0n}^{\left( cl \right)}=T_{0n}^{\left( cl \right)}+V=mv_{0n}^{2}/2+V,$$ are given by the eigenvalues equation ${{\hat{H}}^{\left( cl \right)}}{{\psi }_{n}}=E_{0n}^{\left( cl \right)}{{\psi }_{n}}$ of the classical Hamiltonian operator ${{\hat{H}}^{\left( cl \right)}}$ with respect to the eigenfunctions \eqref{a12} ${{\psi }_{n}}$. It implies that classical Hamiltonian corresponding eigenfunctions \eqref{a12} has been discretized.     Clearly, the complex eigenvalues in this form of energy spectrum correspond to the G-operator \eqref{b6}, it illustrates the G-operator in such an especial condition ${{u}_{x}}=0$ has discrete energy spectrum, it also states that the energy spectrum becomes more complex if ${{u}_{x}}\neq0$ is generally given for the complex eigenvalues. 

According to the eigenvalues  \eqref{d3}, its real part and imaginary part are  respectively shown by
\begin{align}
  & \operatorname{Re}{E_{0n}^{\left( gr \right)}}=E_{0n}^{\left( cl \right)}-{{E}^{\left( s \right)}}\left( {{u_{0}}} \right), \notag\\
 & \operatorname{Im}{E_{0n}^{\left( gr \right)}}=-E_{n}^{\left( q \right)}. \notag
\end{align}
It is found that the form of energy spectrum \eqref{d3} exactly corresponds to the G-operator \eqref{b6} when $u_{x}=0$ is given, this implies that $u_{x}=0$ seems to be a normal condition.  Furthermore, if let $V={{E}^{\left( s \right)}}\left( {{u}_{0}} \right)$, then $E_{0n}^{\left( gr \right)}=mv_{0n}^{2}/2-iE_{n}^{\left( q \right)}$ is the previous result given by \eqref{c1}.

Broadly thinking, the general case is given by \eqref{d1}, therefore, the condition $u_{x}\neq0$ holds for the energy spectrum of the G-operator \eqref{d2},  then it yields the general result
\begin{align}\label{d4}
E_{n}^{\left( gr \right)} &=-{{c}_{1}}\beta _{n}^{\left( II \right)}+V \notag\\
 & =\frac{m}{2}{{v}_{n}}^{2}+\frac{3}{4}{{c}_{1}}\frac{u_{x}^{2}}{{{u}^{2}}}
-\frac{{{c}_{1}}}{2}\frac{{{u}_{xx}}}{u}+2{{c}_{1}}{{u}^{2}}+V
-iE_{n}^{\left( q \right)}\left( 1-\frac{{{u}_{x}}}{{{u}^{2}}} \right),\\
& =E_{n}^{\left( cl \right)}+\frac{3}{4}{{c}_{1}}\frac{u_{x}^{2}}{{{u}^{2}}}
-\frac{{{c}_{1}}}{2}\frac{{{u}_{xx}}}{u}-{{E}^{\left( s \right)}}\left( {{u}} \right)-iE_{n}^{\left( q \right)}\left( 1-\frac{{{u}_{x}}}{{{u}^{2}}} \right),\notag
\end{align}
that corresponds to the eigenfunction \eqref{a12},
where ${{v}_{n}}=w_{n}^{\left( q \right)}/{{u}}$ has been used, and the discrete spectrum of classical kinetic energy  $T_{n}^{\left( cl \right)}=\frac{m}{2}v_{n}^{2}$ with respect to the eigenfunction \eqref{a12}, and $$E_{n}^{\left( cl \right)}= T_{n}^{\left( cl \right)}+V=\frac{m}{2}v_{n}^{2}+V.$$ according to the eigenvalues equation ${{\hat{H}}^{\left( cl \right)}}{{\psi }_{n}}=E_{n}^{\left( cl \right)}{{\psi }_{n}}$ of the classical Hamiltonian operator ${{\hat{H}}^{\left( cl \right)}}$.  As a result, it gets the real part and imaginary part of the eigenvalues of G-operator \eqref{d2}  respectively given by
\begin{align}
  & \operatorname{Re}{E_{n}^{\left( gr \right)}}=E_{n}^{\left( cl \right)}-{{E}^{\left( s \right)}}\left( {{u}} \right)+\frac{3}{4}{{c}_{1}}\frac{u_{x}^{2}}{{{u}^{2}}}
-\frac{{{c}_{1}}}{2}\frac{{{u}_{xx}}}{u}, \notag\\
 & \operatorname{Im}{E_{n}^{\left( gr \right)}}=-E_{n}^{\left( q \right)}\left( 1-\frac{{{u}_{x}}}{{{u}^{2}}} \right)=-E_{n}^{\left( q \right)}+E_{n}^{\left( q \right)}\frac{{{u}_{x}}}{{{u}^{2}}}. \notag
\end{align}Hence, if $u_{x}=0$, then general situation \eqref{d4} is degenerated to the simple case \eqref{d3}.

In general speaking, by using a specific eigenfunctions \eqref{a12} ${{\psi }_{n}}$, when it is acted onto a quantum operator, it always produces the accurate complex eigenvalues, summarily. In this way, we can deduce the complex eigenvalues of the four different kinds of the non-Hermitian Hamiltonian operator, previously. It obtains the complex energy spectrum by studying the
generalized Laplacian equipped with the G-dynamics of the type I and type II, type III, it opens a wider vision to understand how the movement of the quantum particle.

\subsection{On composite solution of generalized Laplacian}
Let's rewrite the one-dimensional eigenvalues equation
of the G-dynamics of the type I in a general form  ${{\hat{w}}^{\left( cl \right)}}{{\psi }}=w^{\left( q \right)}{{\psi }}$,  then we consider the composite solution
 $\Psi =\psi \varphi $ for the generalized Laplacian
on region $\Omega$ \eqref{a13}, where $\psi $ is given by the G-dynamics of type I while
$\varphi$ is another function, it is analogue to the method of separation of variables,  we think of this case, to figure out how it behaves in the generalized Dirichlet Laplacian \eqref{a7}. Namely, the eigenvalues equation
\[\hat{Z}\Psi =-\Delta \Psi +i{{\hat{w}}^{\left( cl \right)}}\Psi ={{\beta }^{\left( 2 \right)}}\Psi, \]of generalized Laplacian with some boundaries, by direct calculation, we obtain
\begin{align}
  & {{\beta }^{\left( 2 \right)}}\psi \varphi =-\varphi \Delta \psi -\psi \Delta \varphi -2\nabla \psi \cdot \nabla \varphi \notag \\
 & \begin{matrix}
   {} & {} & {} & {} & {}& {} & {} \\
\end{matrix}+i\varphi {{{\hat{w}}}^{\left( cl \right)}}\psi +2i{{b}_{c}}\psi \nabla s\cdot \nabla \varphi,  \notag
\end{align}
where if $\varphi$ and $\psi$ are two functions, the Laplace operator of their product is
\[\Delta \left( \psi \varphi  \right)=\varphi \Delta \psi +\psi \Delta \varphi +2\nabla \psi \cdot \nabla \varphi, \]
and
\begin{align}
{{{\hat{w}}}^{\left( cl \right)}}\left( \psi \varphi  \right)  &={{b}_{c}}\left( 2\nabla s\cdot \nabla \left( \psi \varphi  \right)+\psi \varphi \Delta s \right) \notag\\
 & =\varphi {{{\hat{w}}}^{\left( cl \right)}}\psi +2{{b}_{c}}\psi \nabla s\cdot \nabla \varphi  \notag\\
  & =\varphi {{{\hat{w}}}^{\left( cl \right)}}\psi+\psi {{{\hat{w}}}^{\left( cl \right)}}\varphi-{b}_{c} \psi \varphi\Delta s.\notag
\end{align}As a result, it yields
\begin{align}
 {{\beta }^{\left( 2 \right)}}\psi \varphi & =\varphi \hat{Z}\psi -\psi \Delta \varphi -2\nabla \psi \cdot \nabla \varphi +2i{{b}_{c}}\psi \nabla s\cdot \nabla \varphi  \notag\\
 & =\beta \varphi \psi -\psi \Delta \varphi -2\nabla \psi \cdot \nabla \varphi +2i{{b}_{c}}\psi \nabla s\cdot \nabla \varphi.  \notag
\end{align}
Then \[\alpha \varphi \psi =-\psi \Delta \varphi -2\nabla \psi \cdot \nabla \varphi +2i{{b}_{c}}\psi \nabla s\cdot \nabla \varphi, \]where $\hat{Z}\Psi -\varphi \hat{Z}\psi =\alpha \Psi $, and $\alpha ={{\beta }^{\left( 2 \right)}}-\beta$, or \[{{\beta }^{\left( 2 \right)}}\psi \varphi =\varphi \hat{Z}\psi +\psi \hat{Z}\varphi -2\nabla \psi \cdot \nabla \varphi -i{{b}_{c}}\psi \varphi \Delta s,\]where \[2i{{b}_{c}}\psi \nabla s\cdot \nabla \varphi =i\psi {{\hat{w}}^{\left( cl \right)}}\varphi -i{{b}_{c}}\psi \varphi \Delta s,\] it yields \[\left( {{\beta }^{\left( 2 \right)}}+i{{b}_{c}}\Delta s \right)\psi \varphi =\varphi \hat{Z}\psi +\psi \hat{Z}\varphi -2\nabla \psi \cdot \nabla \varphi. \]
It is obvious that we now know the specific expression $\psi$ derived by \eqref{b1}, but another function $\varphi$ is unknown, therefore, the eigenvalues in a certain form are completely determined by the accurate expression of $\varphi$.
For this case that follows, consider
\[\int\limits_{\Omega }{\alpha \varphi \psi d\tau}=-\int\limits_{\Omega }{\psi \Delta \varphi d\tau}-2\int\limits_{\Omega }{\nabla \psi \cdot \nabla \varphi d\tau}+2i{{b}_{c}}\int\limits_{\Omega }{\psi \nabla s\cdot \nabla \varphi d\tau}.\]
More precisely, it has
\begin{align}
 -\alpha \int\limits_{\Omega }{\varphi \psi d\tau} &=\int\limits_{\Omega }{\psi \Delta \varphi d\tau}+2\int\limits_{\Omega }{\nabla \psi \cdot \nabla \varphi d\tau}-2i{{b}_{c}}\int\limits_{\Omega }{\psi \nabla s\cdot \nabla \varphi d\tau} \notag\\
 & =\int\limits_{\Omega }{\nabla \varphi \cdot \nabla \psi d\tau}+\int\limits_{\partial \Omega }{\varphi \frac{\partial \psi }{\partial \boldsymbol {n}}dS\left( x \right)}-2i{{b}_{c}}\int\limits_{\Omega }{\psi \nabla s\cdot \nabla \varphi d\tau} \notag\\
 & =\int\limits_{\Omega }{\nabla \varphi \cdot \nabla \psi d\tau}-2i{{b}_{c}}\int\limits_{\Omega }{\psi \nabla s\cdot \nabla \varphi d\tau}. \notag
\end{align}
Using the formula \[2i{{b}_{c}}\int\limits_{\Omega }{\psi \nabla s\cdot \nabla \varphi d\tau}=i\int\limits_{\Omega }{\left( {{{\hat{w}}}^{\left( cl \right)}}\varphi -{{b}_{c}}\varphi \Delta s \right)\psi d\tau},\]we get the further derivation
\begin{align}
 -\alpha \int\limits_{\Omega }{\varphi \psi d\tau} & =\int\limits_{\Omega }{\nabla \varphi \cdot \nabla \psi d\tau}-i\int\limits_{\Omega }{\left( {{{\hat{w}}}^{\left( cl \right)}}\varphi -{{b}_{c}}\varphi \Delta s \right)\psi d\tau} \notag\\
 & =\int\limits_{\Omega }{\nabla \varphi \cdot \nabla \psi d\tau}-i{{b}_{c}}\int\limits_{\Omega }{\psi \left( \nabla s\cdot \nabla \varphi  \right)d\tau} \notag\\
 & \begin{matrix}
   {} & {} & {}  \\
\end{matrix}-i{{b}_{c}}\int\limits_{\Omega }{\varphi \left( \nabla s\cdot \nabla \psi  \right)d\tau}-i\int\limits_{\Omega }{\left( {{{\hat{w}}}^{\left( cl \right)}}\varphi  \right)\psi d\tau} .\notag
\end{align}In this way, it can provide some clues to the properties of the  composite solution.
What if $\varphi =\psi$ is given, then
\[2i{{b}_{c}}\int\limits_{\Omega }{\psi \left( \nabla s\cdot \nabla \psi  \right)d\tau}=\int\limits_{\Omega }{{{\left| \nabla \psi  \right|}^{2}}d\tau}+\alpha \int\limits_{\Omega }{{{\psi }^{2}}d\tau}\geq\alpha \int\limits_{\Omega }{{{\psi }^{2}}d\tau}.\]
As the case of the composite solution performs, the situation becomes more complex, because of we have to consider the certain expression of the functions $\psi$ and $\varphi$ at the same time. As a consequence, we need to deeply seek more clues to solve it. It needs to be emphasized that the eigenvalue equations of eigenfunctions
both given by the G-dynamics of type I as a imaginary part of generalized Laplacian are clear to see and use, this usefully provides us with a specific direction to study generalized Laplacian equipped with different boundary condition in using the eigenfunctions deduced by the G-dynamics of type I just like we do in one-dimensional case of generalized Laplacian.

\section{Conclusions}

On the foundation of the eigenvalue problem of Laplace operator, the
different boundary conditions are given to certain discussions such as
Dirichlet boundary condition, Neumann boundary condition, etc. The study all shows the similar results associated with the discrete spectrum of infinitely many non-negative eigenvalues with no finite accumulation.
By analogy, the differential model of the generalized Laplacian equipped with the G-dynamics of the type I abstractly comes from the motor operator which depicts the quantum energy spectrum, it can help to understand how the energy spectrum is distributed, in particular, the G-dynamics of the type I shows us more peculiar features such as the integral identity. Then type II for generalized Laplacian as a standard second order differential equation has been considered, especially, in  one-dimensional eigenvalues equation. At last, we take the composite solution of generalized Laplacian into discussions under the condition of one eigenfunction certainly given by the G-dynamics of the type I.  We have explained that the energy spectrum can be corresponded to the classical menchanics if the eigenfunction certainly given by the G-dynamics of the type I is considered. Meanwhile, the type II and type III for generalized Laplacian are together analyzed.

\section*{References}
\ \ \
\par [1]
Gottlieb HP. Eigenvalues of the Laplacian with Neumann boundary conditions. The ANZIAM Journal. 1985, 26(3):293-309.
\par [2] Hile GN, Protter MH. Inequalities for eigenvalues of the Laplacian. Indiana University Mathematics Journal. 1980, 29(4):523-38.
\par [3]
McKean Jr HP, Singer IM. Curvature and the eigenvalues of the Laplacian. Journal of Differential Geometry. 1967, 1(1-2):43-69.

\par [4] Arfken G. Self-Adjoint Differential Equations. in Mathematical Methods for Physicists, 3rd ed. Orlando, FL: Academic Press, 1985, 497-509.
\par [5]
Kuttler JR, Sigillito VG. Eigenvalues of the Laplacian in two dimensions. Siam Review. 1984, 26(2):163-93.

\par [6]
Moretti V. Spectral Theory and Quantum Mechanics:  Mathematical Foundations of Quantum Theories, Symmetries and Introduction to the Algebraic Formulation, Springer-Verlag, 2018.
\par
 [7] Whongius J. A rigorous Hermitian proof about the G-dynamics and analogy with Berry-Keating's Hamiltonian.  arXiv:2109.03068.
\par [8]
Wolf SA, Keller JB. Range of the first two eigenvalues of the Laplacian. Proceedings of the Royal Society of London. Series A: Mathematical and Physical Sciences. 1994, 447(1930):397-412.
\par [9]
	Weyl H. Ramifications, old and new, of the eigenvalue problem Bull. Amer. Math. Soc. 1950, 56: 115–139.
\par [10]
Weyl H. Das asymptotische Verteilungsgesetz der Eigenwerte linearer partieller Differentialgleichungen Math. Ann. 1911, 71: 441–479.
\par [11]
Henrot, Antoine. Minimization problems for eigenvalues of the Laplacian. Nonlinear Evolution Equations and Related Topics. Birkhäuser, Basel, 2003. 443-461.
\par [12]Ashbaugh, Mark S. Open problems on eigenvalues of the Laplacian. Analytic and geometric inequalities and applications. Springer, Dordrecht, 1999, 13-28.

\par [13]
Pockels F. Über die partielle Differentialgleichung $\Delta \text{u+}{{\text{k}}^{\text{2}}}\text{u=0}$ und deren Auftreten in die mathematischen Physik Z. Math. Physik , 1892, 37:100–105

\par [14]
Cheng QM, Yang H. Bounds on eigenvalues of Dirichlet Laplacian. Mathematische Annalen. 2007, 337(1):159-75.

\par [15]
Kroger P. Estimates for sums of eigenvalues of the Laplacian. Journal of Functional Analysis. 1994,126(1):217-27.

\par [16]
Gilkey PB. Curvature and the eigenvalues of the Laplacian for elliptic complexes. Advances in Mathematics. 1973, 10(3):344-82.

\par [17]
Tanno S. Eigenvalues of the Laplacian of Riemannian manifolds. Tohoku Mathematical Journal, Second Series. 1973, 25(3):391-403.
\par [18]
 Laptev A. Dirichlet and Neumann eigenvalue problems on domains in Euclidean spaces. journal of functional analysis. 1997,151(2):531-45.

\end{document}